\documentclass[aps,pra,twocolumn,preprintnumbers,amsmath,footinbib,10pt,longbibliography]{revtex4-2}
\pdfoutput=1
\makeatletter
\renewcommand*{\@fnsymbol}[1]{\ensuremath{\ifcase#1\or \or \spin\or \covbond\or \boson\or \orbit\or \tachyon\or \magnon\or \atom\or \*\or \quadrupole\or
   \mathsection\or \mathparagraph\or \|\or **\or \dagger\dagger
   \or \ddagger\ddagger \else\@ctrerr\fi}}
\makeatother

\usepackage{hyperref}
\usepackage[utf8]{inputenc}
\usepackage[sort&compress]{natbib}
\usepackage[dvipsnames]{xcolor}
\usepackage{graphicx}
\usepackage{epsfig}
\usepackage{epstopdf}
\usepackage{amsfonts}
\usepackage{amstext,amsmath,amssymb}
\usepackage{svrsymbols}
\usepackage{wasysym}

\usepackage{rotating}
\usepackage{dcolumn}
\usepackage{array,multirow}
\usepackage[english]{babel}
\usepackage[export]{adjustbox}
\usepackage{braket}
\usepackage[caption=false]{subfig}
\usepackage{xr}
\usepackage{ifpdf}
\usepackage{tikz}

\usetikzlibrary{arrows}

\hypersetup{urlcolor=violet, citecolor=violet, linkcolor=violet, colorlinks=true}
\newcommand{\be}{\begin{equation}}
\newcommand{\ee}{\end{equation}}
\newcommand{\bea}{\begin{eqnarray}}
\newcommand{\eea}{\end{eqnarray}}

\newcommand{\round}[1]{\ensuremath{\left\lfloor#1\right\rceil}}

\usepackage{color}

\def\brms{B_\text{rms}}
\def\frms{\Phi_\text{rms}}

\DeclareMathOperator{\sinc}{sinc}

\emergencystretch=\maxdimen
\begin{document}

\title{Optimal Sensing Protocol for Statistically Polarized Nano-NMR with NV Centers}

\author{Nicolas Staudenmaier${}^{1,\tachyon}$, Anjusha Vijayakumar-Sreeja${}^{1,\tachyon}$, Genko Genov${}^{1,\tachyon}$, Daniel Cohen${}^{2}$, Christoph Findler${}^{3}$, Johannes Lang${}^{3}$, Alex Retzker${}^{2,4}$, Fedor Jelezko${}^{1}$, Santiago Oviedo-Casado${}^{2,5,\orbit}$}

\affiliation{${}^{1}$Institute for Quantum Optics, Ulm University, Albert-Einstein-Allee 11, 89081 Ulm, Germany\\
${}^{2}$Racah Institute of Physics, Hebrew University of Jerusalem, 91904 Jerusalem, Israel\\
${}^{3}$Diatope GmbH, Buchenweg 23, 88444 Ummendorf, Germany\\
${}^{4}$AWS Center for Quantum Computing, Pasadena, California, USA\\
${}^{5}$\'Area de F\'isica Aplicada, Universidad Polit\'ecnica de Cartagena, Cartagena E-30202, Spain
}

\thanks{${}^{\tachyon}$ This authors contributed equally}
\email{\\${}^{\orbit}$oviedo.cs@mail.huji.ac.il}

\begin{abstract}
 Diffusion noise represents a major constraint to successful liquid state nano-NMR spectroscopy. Using the Fisher information as a faithful measure, we theoretically calculate and experimentally show that phase sensitive protocols are superior in most experimental scenarios, as they maximize information extraction from correlations in the sample. We derive the optimal experimental parameters for quantum heterodyne detection (Qdyne) and present the most accurate statistically polarized nano-NMR Qdyne detection experiments to date, leading the way to resolve chemical shifts and $J$ couplings at the nanoscale.
\end{abstract}


\maketitle

Phase sensitive protocols, such as quantum heterodyne detection (Qdyne),
optimize the precision of frequency estimation from classical signals in noisy environments by sequentially sampling the probe at periodic intervals, 
thereby overcoming the limited coherence time of the quantum sensor \cite{Schmitt2017,Degen2017,Glenn2018}. Then, resolving arbitrarily close frequencies should be possible by increasing the total measurement time \cite{Rotem2019,Gefen2019}. These results heralded the possibility of performing effective nuclear magnetic resonance at the nanoscale (nano-NMR), on statistically polarized liquid samples, with quantum probes; for example, single nitrogen-vacancy (NV) centers in diamond \cite{Staudacher2013,Mamin2013,Muller2014,Ajoi2015,Lovchinsky2016}. However, despite the early promising results, these protocols have been scarcely applied in experiments with chemically or pharmacologically relevant samples \cite{Aslam2017,Glenn2018,Bucher2020,Farida2021,BucherLOC,Freire-Moschovitis2023,Nie2021}. The foremost difficulty explaining this reluctance is the necessary trade-off between sensor-sample interaction strength and short diffusion time, typical of statistically polarized samples. This combines with a seemingly challenging data acquisition and postprocessing, and the existence of more amenable alternatives such as correlation spectroscopy (CS) \cite{Laraoui2013,Staudacher2013}.

The archetypal color defects used for single NV nano-NMR are located at depths $d$ a few nm below the diamond surface, where they interact via dipole-dipole coupling with nuclei from a sample located on top of the diamond surface. With the interaction strength decreasing as $\sim d^{-3}$, only the closest nuclei contribute significantly to the total interaction. Then, for shallow NV centers, the statistical polarization of the nuclei is significant enough to overcome any thermal averaging \cite{Degen2007,Reinhard2012,Staudacher2013,Herzog2014}. These statistically polarized nuclei generate a time correlated magnetic field 
$B(t) = \sum_i a_i(t)\cos(\omega_i t) + b_i(t)\sin(\omega_i t)$, that can contain oscillations at several (e.g. Larmor) frequencies $\omega_i$. $B(t)$ generates a detectable change on the NV center electron spin state, thereby providing valuable information about the sample. However, as molecules diffuse, the couplings between nuclei and NV center ---the amplitudes $\{a(t),b(t)\}$--- fluctuate. This is typically modeled as a random process normally distributed around zero and with finite correlation time $T_D = d^2/D$, with $D$ the diffusion coefficient of the fluid \cite{Pham2016}. It is commonly accepted that, when the diffusion time $T_D$ is shorter than an oscillation period of $B(t)$, one cannot accumulate enough information per measurement to resolve the respective spectral line. Yet recently it was demonstrated that correlations between diffusing nuclei survive longer than $T_D$, allowing to significantly extend the data acquisition time and, consequently, the precision of estimation of the target frequency \cite{Cohen2020,Oviedo2020,TuviaUltimate,Staudenmaier2022}.

In this Letter, we determine the optimal experimental parameters for two state-of-the-art quantum sensing protocols, aiming to get the maximum information for spectral reconstruction. We combine this strong theoretical foundation with robust data analysis, allowing us to present the most accurate statistically polarized nano-NMR Qdyne experiments to date. We compare them to ideal, error-free CS experiments through the amount of information that they provide about a target signal frequency, and unambiguously establish that Qdyne has a superior performance. These results open new possibilities for high precision quantum sensing, broadening the scope for implementing Qdyne in liquid nano-NMR, conceivably allowing to resolve chemical shifts and $J$ couplings at the nanoscale. Additionally, we demonstrate a universal comparison methodology for sensing protocols and experimental platforms that can potentially become a valuable tool for optimizing any quantum sensing experiment.

\textit{Quantifying information.}--- We consider an NV center interacting with the magnetic field $B(t)$ generated by a statistically polarized sample of diffusing nuclei. 
Assuming that any other external noise is strongly suppressed by either dynamical decoupling (DD) \cite{Viola1999,Cywinski2007,KDDref,KDDref2,CasanovaPRA2015,Suter2016,GenovPRL2017} or careful experimental design, the main limiting factor to spectral resolution is the finite correlation time, $T_D$, of $B(t)$. An initial superposition state of the NV center interacting with $B(t)$, accumulates a phase $\Phi[B(t)]$. Tracking the evolution of such state permits inferring the parameters describing $B(t)$ using suitable postprocessing. 

Experimentally, the mean squared error (MSE) $\Delta\delta$ quantifies the accuracy in estimating a parameter; e.g., the small (angular) frequency offset $\delta$, defined as the difference between the nuclear spin Larmor frequency and the sampling frequency (see Appendix~\ref{AppendixB}). The Cram\'er-Rao bound states that the MSE must always be greater than the inverse Fisher information about the parameter, $\Delta\delta \geq 1/I_\delta$ \cite{Wootters1981,Braunstein1994}, establishing a connection between theory and experiments, and providing a direct route to improve experiments through theoretical modeling. For a single measurement 
\begin{equation}
i_\delta = {\mathbb{E}}\left[\left(\frac{d\log[L(\delta)]}{d\delta}\right)^2\right],\label{singleFI}
\end{equation}
with $L(\delta)$ the likelihood of finding the NV in a given state. After $N$ measurements in an experiment of duration $T$, the total information is 
$I_\delta = \sum_N i_\delta$. We aim to theoretically maximize $I_\delta$ for a given protocol, seeking the best possible experimental $\Delta\delta$.

Here, we compare Qdyne with one of the most advanced sensing protocols: correlation spectroscopy \cite{Laraoui2013,Staudacher2013}. We benchmark the achievable $I_\delta$ for each protocol, in the scenario of a magnetic signal with a limited coherence time originating from a statistically polarized sample, by defining the ratio \cite{oviedocasado2022}
\be
R_\delta = I_\delta^\mathrm{Qdyne}/I_\delta^\mathrm{CS}
\label{exactratio}
\ee 
between the Fisher information of Qdyne $I_\delta^\mathrm{Qdyne}$ and CS $I_\delta^\mathrm{CS}$. 
In what follows, we theoretically calculate $I_\delta$ for each protocol showing that, in most experimental scenarios, $R_\delta > 1$. Then, we compare it to $\Delta\delta$ obtained from statistically polarized nano-NMR experiments and demonstrate strong accord with the theoretical modeling.

\begin{figure}
     \includegraphics[width=\columnwidth,height=6cm]{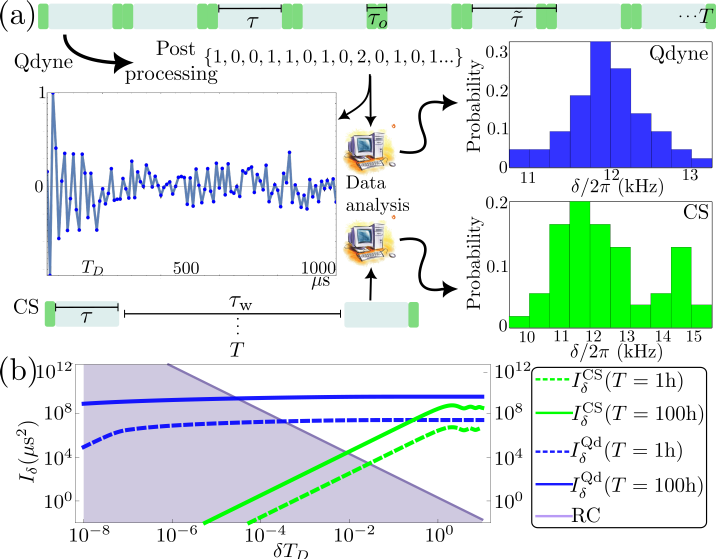}
      \caption{(a) 
      Depiction of Qdyne (Qd, upper part) and correlation spectroscopy (CS, lower part) experimental procedures for NV-NMR on a fluid sample with diffusion time $T_D$. For Qdyne, the measurement sequence is repeated at fixed time intervals $\tilde\tau$. During each overhead time $\tau_\mathrm{o}$ the NV is interrogated about the phase acquired during the time $\tau$ and repolarized for a new measurement. The recorded photons are postselected to obtain a measurement vector whose autocorrelation resembles the covariance of the noise model. In CS, the autocorrelation of the sample is probed directly by changing the waiting $\tau_\mathrm{w}$ between two  DD blocks. Data analysis using maximum likelihood estimation produces a histogram of estimators which, for low frequencies, is narrower for Qdyne than for CS. (b) Exact Fisher information $I_\delta$ for both CS [Eq.~(\ref{cstotalinfo})] and Qdyne [Eq.~(\ref{qdynetotalinfo})] with total measurement time $T$ = 1 hour and $T$ = 100 hours. The shadowed area marks the frequencies for which the information is insufficient for successful estimation, set at $\Delta\delta < \delta^2/4$, as per the Rayleigh criterion (RC). For all $I_\delta$ $\frms = 1$, with $T_D$ = 100\,$\mu$s, $\eta_0 = 0.04$, $\eta_1 = 0.03$, and for Qdyne $\tilde\tau$ = 25\,$\mu$s.}\label{Fig1}
\end{figure}

\textit{Correlation spectroscopy.}--- 
Standard quantum spectroscopy with NV centers monitor their fluorescence response, which is modulated by the accumulated phase $\Phi[B(t)]$ in each measurement. Then, one infers information about $B(t)$ by averaging the results of all measurements. Since $B(t)$ is time correlated, consecutive phase acquisition periods are also correlated, and encode information about $B(t)$. CS builds upon this notion by combining two phase acquisition periods of equal duration $\tau$ separated by a waiting time $\tau_\mathrm{w}$ [see Fig.\ref{Fig1}(a)]. The phase accumulated during the first period, is stored as population difference (transferred to a ``memory'' qubit) during $\tau_\text{w}$ \cite{Laraoui2013}, imposing $\tau_\mathrm{w} \leq T_1$ ($T_{m}$), the latter being the spin-lattice (memory qubit) relaxation time. Varying $\tau_{\mathrm{w}}$ yields a signal that corresponds to the covariance of the noise affecting the probe, i.e. ${\text{cov}}\langle\Phi[B(t^\prime)]\Phi[B(t^\prime+t)]\rangle = \Phi_{\text{rms}}^2\cos(\delta t)C(t/T_D)$, where $t=\tau+\tau_\mathrm{w}$, $\frms \propto \brms$ the root-mean-squared field of $B(t)$, and $C(t/T_D)$ is the envelope that describes the decay of correlations due to noise. For diffusion, $C(t/T_D \ll 1) \sim \exp(-6t/T_D)$ while $C(t/T_D \gg 1) \sim t^{-3/2}$ \cite{Cohen2020}.

In each readout, the expected photon number is $p = \eta + c\langle \sin(\Phi_0)\sin(\Phi_t)\rangle/2$, with $\eta = (\eta_0 + \eta_1)/2$ the average photon number ($\eta_0$ and $\eta_1$ are the expected photon numbers for readout of NV spin states $\ket{0}$ and $\ket{1}$, respectively), and $c = \eta_0 - \eta_1$ the contrast. Averaging over realizations of $B(t)$ yields the relation between the expected photon number and the autocorrelation, $p = \eta + c\,\frms^2\cos(\delta t)C(t/T_D)/2$. Using Eq.~(\ref{singleFI}) with $p$ we get $i_\delta^\mathrm{CS} = \frac{c^2}{4\eta + c^2}\frms^4t^2\sin^2(\delta t)C^2(t/T_D)$. To calculate $I_\delta$, we transform the measurements sum into an integral, and take the low frequency limit $(\delta/2\pi) < 1/T_D$ (see Appendix~\ref{AppendixQdyneInfo} for details). Then \be
I_\delta^\mathrm{CS} =  TT_D\int_0^1 dt\, i_\delta^\mathrm{CS}(t) \approx \frac{c^2\frms^4\delta^2 T_D^3T}{4\eta + c^2},  \label{cstotalinfo}
\ee
where we assume $T_1 (T_{m}) \gg T_D$ and ignore its effect. The relation between $\delta$, $T_D$, and $T$ defines the ability to estimate $\delta$.
If $(\delta/2\pi) T_D > 1$, a complete oscillation occurs before the signal is strongly suppressed, and estimating the frequency poses no problem. Conversely, for $(\delta/2\pi) T_D < 1$, $\lim_{\delta \rightarrow 0} I_\delta^\mathrm{CS}(T) = 0$. 
This is illustrated in Fig.~\ref{Fig1}(b), where we calculate $I_\delta^\mathrm{CS}(T)$ using the exact analytical formula, for two different total experiment times T, and compare it against the maximum MSE that allows for frequency estimation, defined, following the Rayleigh criterion for linewidth separation in optics, as $\Delta\delta < \delta^2/4$ \cite{Abbe1873,Rayleigh1879,JonesAR1995}. We observe that $I_\delta$ vanishes with $\delta$ faster than it grows with $T$. Note that both the oscillations and the relative flatness appearing at $(\delta/2\pi) T_D \gtrapprox 1$ result from using the exact $I_\delta^\mathrm{CS}$ rather than the approximate expression Eq.~(\ref{cstotalinfo}).

\textit{Qdyne.}--- Phase sensitive measurements consist of (phase-dependent) signal accumulation periods $\tau$ followed by qubit interrogation and repolarization (for overhead time $\tau_\mathrm{o}$) for the next measurement, in a process repeated continuously with periodic separation time $\tilde\tau = \tau + \tau_\mathrm{o}$, as shown in Fig.~\ref{Fig1}(a) (see, also, Appendix~\ref{Appendix:Qdyne} for a detailed description of Qdyne implementation). Following each interrogation, the expected photon number is $p = \eta + c\langle \sin(\Phi_t)\rangle/2$. A sequence of such equally spaced measurements directly reflects the time evolution of $B(t)$. To better understand Qdyne, we use polar coordinates and write $B(t) = \Omega(t)\cos\left[\delta t + \varphi(t)\right]$, with $\Omega(t) = \sqrt{a(t)^2 + b(t)^2}$ and $\tan\left[\varphi(t)\right] = b(t) / a(t)$. Then, when $T_D \gg 1/\delta$, $\tilde\tau \ll T_D$ and $\varphi(t)$ remains coherent for a sufficiently large number of measurements, allowing us to perform maximum likelihood estimation (MLE) over a small parameter space. Increasing $T$ permits resolving arbitrarily small frequencies \cite{Rotem2019}.

When $T_D < 1/\delta$, $\varphi(t)$ changes faster than the oscillation period $2\pi/\delta$. Then, each new measurement requires estimating new parameters, and MLE becomes computationally intractable. In this scenario, the solution is to calculate the autocorrelation of the measured signal during postprocessing [see Fig.~\ref{Fig1}(a)]. This treatment reduces the parameter space by having a decaying signal similar to that of CS. Then, the signal parameters can be estimated through, e.g., least squares algorithms. The advantage of Qdyne is that, by measuring sequentially, any two points in a measurement vector are correlated, instead of having just pairwise correlation, as in CS. For the power-law signals occurring in nanoscale diffusion \cite{Cohen2020,Staudenmaier2022}, this means that much more information is contained in the Qdyne autocorrelation. The relevant probability is that of obtaining a pair of correlated photons between any two measurements $\langle p_0p_{j\tilde\tau}\rangle$, $j$ being an integer, yielding $i_{\delta}^\mathrm{Qd} = \frac{c^4}{\left(4\eta + c^2 \right)^2} \frms^4 t^2\cos(\delta t)C^2(t/T_D)$. We obtain the total information by integrating over $t$, with an added factor accounting for all correlated pairs of measurements (details of the calculation can be found in Appendix~\ref{AppendixQdyneInfo}). Using normalized time $z = t/T_D$ we get \be
I_\delta^\mathrm{Qdyne} =  \frac{T_D^4}{\tilde\tau^2}\int_0^{\frac{T}{T_D}} dz \,i_\delta^\mathrm{Qd}\left(\frac{T}{T_D}-z\right) \approx \frac{c^4\frms^4T_D^3T\log{\delta T}}{\left(4\eta + c^2 \right)^2\tilde\tau^2}, \label{qdynetotalinfo}
 \ee  where we consider $(\delta/2\pi) T_D < 1$ and $(\delta/2\pi) T > 1$. Equation~(\ref{qdynetotalinfo}) shows that, with Qdyne, spectral resolution no longer depends on the relation between $T_D$ and $\delta$, but solely on $T$, as illustrated in Fig.~\ref{Fig1}. There, only when $(\delta/2\pi) T < 1$ does $I_\delta^\mathrm{Qdyne} \rightarrow 0$. 
 
Equation~(\ref{qdynetotalinfo}) has several implications: Qdyne requires a high number of measurements, and therefore a small $\tau_\mathrm{o} = \tilde\tau-\tau$, and two measurements per correlation (whereas CS requires one), which translates to an extra $\frac{c^2}{4\eta + c^2}$, penalizing Qdyne due to the typically low average number of photons detected per measurement. Conversely, a small $(\delta/2\pi) T_D$ favors Qdyne as compared to CS. Thus, increasing the measurement time $T$ permits estimating much smaller frequencies. The information ratio, Eq.~(\ref{exactratio}), in the limit $\delta/2\pi < 1/T_D$ reads 
 \be
 R_\delta = \frac{I_\delta^\mathrm{Qdyne}}{I_\delta^\mathrm{CS}}\approx \frac{c^2}{4\eta + c^2}\frac{\log(\delta T)}{\delta^2\tilde\tau^2}.
\ee
Next, we theoretically explore the dependence of $R_\delta$ on different parameters and compare it to experimental results.



\textit{Experimental results and data analysis.}--- To test our theoretical model, we perform nano-NMR measurements on statistically polarized samples with single shallow NV centers. For both protocols we record the signal coming from hydrogen nuclei in an immersion oil (Fluka 10976) with a Larmor frequency $\approx 2\,\mathrm{MHz}$ (see Appendix~\ref{AppendixA} and Ref.~\cite{Staudenmaier2022} for further details).

\begin{figure*}
\includegraphics[width=16cm,height=4.8cm]{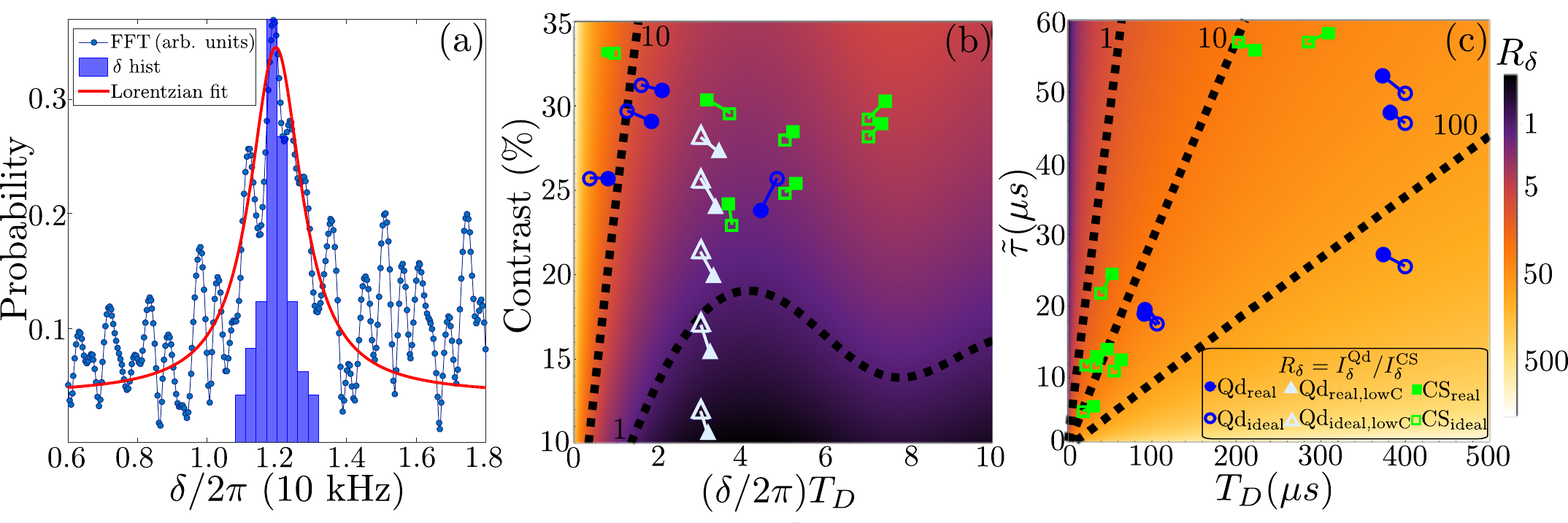}
    \caption{Probability histogram of frequency estimators from several autocorrelation slices of a Qdyne (Qd) experiment and the Fourier transform of the full autocorrelation in (a). In (b) and (c), exact information ratio from Eq.~(\ref{exactratio}) in logarithmic scale in the background. Ratio between the information obtained experimentally and the ideal information for the same parameters with the alternative protocol. Empty shapes represent the theoretical expectation for $R_\delta$ while full shapes show the experimental result. 
    Thick dashed lines show reference $R_\delta$. In (b), $T$ = 100\,hours, $\tilde\tau = 40\,\mu$\,s, and $T_D = 50\,\mu$\,s, while for (c) $(\delta/2\pi) T_D = 1$, with constant overhead time $\tau_\mathrm{o} = 2.1\,\mathrm{\mu s}$ \cite{Degen2017}. Note that we use relative contrast defined as $(\eta_0-\eta_1)/\eta_0$, for which purpose we scale Eqs.~(\ref{cstotalinfo}) and (\ref{qdynetotalinfo}) appropriately and we fix it to 25\,\% ($\eta_0 = 0.04$, $\eta_1=0.03$) in (c). The legend and color bar in (c) apply to (b) as well.} \label{Experiments}
\end{figure*}

Data postprocessing is important for successful nano-NMR, especially when $(\delta/2\pi) T_D < 1$. Minimizing estimation errors requires using efficient estimators such as MLE \cite{Dekking2005}, that can saturate the Cram\'er-Rao bound. Contrary to that, the typically employed Fourier transform offers sufficient statistics for parameter estimation only for signals with well separated frequencies and high signal-to-noise ratio. Moreover, estimating frequencies from Fourier peaks is sub-optimal \cite{Bretthorst1988}, explaining its failure in the challenging nano-NMR regimes, which feature highly fluctuating coupling parameters and for frequency splittings that are smaller than $1/T_D$.

Our goal is to optimize $\Delta \delta$ and compare it to the theoretical calculations of $I_\delta$. To have the most faithful comparison we need MLE. However, as noted above, Qdyne measurement vectors become computationally intractable for MLE when $(\delta/2\pi) T_D < 1$. Rather, we resort to fitting the autocorrelation of the data to $\Phi_{\text{rms}}^2\cos(\delta t)C(t/T_D)$ with parameters $\{\frms,\delta,T_D\}$ using a simple nonlinear least squares algorithm, which is equivalent to MLE if the variables to be fitted have normally distributed errors (see Appendix~\ref{MLE} and Ref.~\cite{ANH1988317} for details), which we can assume. Estimating the MSE from a single fit to the autocorrelation can be misleading, especially in the hardly resolvable regime, due to outliers, a highly nonconcave parameter space, and because a good initial estimation of the MSE is required. Instead, we calculate the root MSE (rmse) from a distribution of frequency estimators, obtained by fitting smaller blocks of the data. These are calculated by slicing the measurement vector onto 15 minute ($\sim 10^7$ measurements) pieces, and then combining them in groups of 20. Thus, we achieve a balance between reducing noise and minimizing statistical errors. We benchmark the acquired squared inverse of the rmse against $I_\delta$. A demonstration of the difference in data analysis procedures is shown in Fig.~\ref{Experiments}(a), where we display the probability histogram of $\delta$ estimators, with rmse = 458 Hz, together with the Fourier transform of the autocorrelation of the entire measurement vector, with rmse = FWHM/2 = 602 Hz \cite{Bretthorst1988,Schmitt2021}. To the best of our knowledge, the former represents the most precise statistically polarized Qdyne result to date \cite{Schmitt2017}. 

In correlation spectroscopy, each experiment yields a single autocorrelation of $\Phi[B(t)]$,
which we fit with a nonlinear least squares algorithm. We evaluate the data 100 times to the model with random initial parameters and take the best fit, in terms of $R^2$. The frequency estimator and its rmse are obtained from the parameter estimates from the best model.

To obtain $R_\delta$ we calculate, for each experiment, the theoretical $I_\delta$ for the alternative measurement protocol using the same parameters as in the experiment. We resort to this procedure given the difficulty of performing identical experiments with both protocols. This procedure excludes possible error sources present in the experiments, which results in a $R_\delta$ biased toward the theoretical calculation, in which these same errors are excluded. This accounts, in part, for not saturating the Cram\'er-Rao bound (see Appendix~\ref{CR}).

Figures \ref{Experiments}(b) and (c) show the experimental results for $R_\delta$ over a theoretical background calculated exactly from the integrals in Eqs.~(\ref{cstotalinfo}) and (\ref{qdynetotalinfo}), fixing all the parameters but two. This poses a problem when showing together theory and experiment, as different experiments have more than two different parameters. We solve the issue by scaling the experimental results according to the fixed parameters chosen for the theory background. This does not alter the relation between protocols (the experimental $R_\delta$ still reflects the best protocol for these parameters), and permits showing the optimal parameter regions for each protocol while at the same time demonstrating correspondence between theory and experiments. In each figure, we display the obtained ratio with the corresponding experimental result as filled shapes in the parameter space, while empty shapes represent full theoretical expectation. 
There is a small deviation between the full theoretical expectation and the ratio between theory and experiment. The similarity between the deviations for CS and Qdyne experiments show that both experimental setups are subject to similar, not-accounted-for, errors (see Appendix~\ref{CR}). This represents a strong validation of our theoretical modeling, and shows that calculating $I_\delta$ is a mathematically consistent procedure for experimental optimization.

The main challenge for statistically polarized nano-NMR is to resolve spectral lines whose frequency difference is smaller than the inverse characteristic time $1/T_D$. In Fig.~\ref{Experiments}(b) we compare the performance of Qdyne and CS according to the relation between the target frequency and $T_D$. For small $\delta$, the superior ability of Qdyne to capitalize on the long-lived power-law correlations means that it is the protocol of choice. Considering a reasonable $T = 100$\,hours experiment, even at poor contrast, it proves superior to CS.A reduction in the total measurement time to $T=1$\,hour does not significantly change the theoretical superiority of Qdyne, as observed in Fig.~\ref{Fig1}(b). 
We note that long experimental times may be challenging for some biological applications. They can be further reduced by improving photon collection efficiency through engineering the diamond geometry \cite{hadden2010strongly,siyushev2010monolithic,Robledo2011Nature,Momanzadeh2015}, optimizing the sampled times, as specified in Appendix~\ref{Appendix:Qdyne}, or through data postprocessing using machine learning \cite{Aharon2019scirep}. When $(\delta/2\pi) T_D > 1$, the influence of the total measurement time diminishes and the contrast gains importance, with CS preferred at poor contrast. However, typical experiments with single NV centers easily feature relative contrasts $> 20\%$ (rather, in excess of $30\%$), for which Qdyne performs better. To experimentally demonstrate that at low contrast $R_\delta < 1$, we have artificially altered the contrast during the postprocessing of the Qdyne data by choosing the photons from different lengths of the readout window (see Appendix~\ref{Appendix:Qdyne} [triangles in Fig.~\ref{Experiments}(b)].

A crucial factor when setting up a nano-NMR experiment is that $\frms \lesssim 1$ for optimal information acquisition (see Appendix~\ref{Appendix:Qdyne} and \cite{Oviedo2020}). But $\frms \propto \brms\tau$, where the $\brms$ is given by the sample. Then, $\tau$ is relatively fixed by said condition. For CS the effect is negligible, but for Qdyne, whose strategy is to measure at a fast rate to accumulate correlations, longer sequence times $\tilde\tau$ negatively affect $I_\delta^\mathrm{Qdyne}$ [note the  $1/\tilde\tau^2$ factor in Eq.~(\ref{qdynetotalinfo})]. In Fig.~\ref{Experiments}(c), we explore the influence of $\tilde\tau$ on the performance of Qdyne. For longer $T_D$ more information can be sampled with Qdyne than with CS, even for larger $\tilde\tau$. This is reflected by the lines of constant $R_\delta$ with approximately constant slope. Conversely, short diffusion times, resulting from very shallow NV centers, favor CS. But this is usually compensated by the shorter measurement times for NVs at low depth, resulting in Qdyne being superior for most of the parameter range. We note that the total information could be insufficient for parameter estimation for very low $T_D$, regardless of the protocol, as demonstrated in Fig.~\ref{Fig1}.

\textit{Conclusions.}--- Diffusion noise poses a major challenge for liquid nano-NMR, and requires refining the experimental design and data analysis for maximal information extraction. Here, we theoretically show theoretically and experimentally demonstrate that phase-sensitive, sequential measurements, such as Qdyne, combined with maximum likelihood estimation, can significantly enhance frequency estimation precision. We apply them in statistically polarized samples and compare with other advanced techniques such as correlation spectroscopy. 
We present a systematic procedure, based on the Fisher information, that allows us to both compare the protocols and optimize the 
sensing experiments. Our work is applicable to a wide variety of different platforms by changing the respective parameters to reflect the particulars of the experiment.
Accordingly, we managed to perform the most accurate Qdyne experiments in liquid nano-NMR to date, providing a recipe toward executing efficient single NV  nano-NMR with samples of biochemical importance and resolving chemical shifts or measuring $J$ couplings.

\begin{acknowledgments} 
\textit{Acknowledgments.}--- S.O.C. acknowledges support from the Fundaci\'on Ram\'on Areces Postdoctoral Fellowship (XXXI edition of grants for Postgraduate Studies in Life and Matter Sciences in Foreign Universities and Research Centres) and the María Zambrano Fellowship. A.V.S. acknowledges support from the European Union’s Horizon 2020 Research and Innovation Program under the Marie Skłodowska-Curie Grant Agreement No. 766402. N.S. acknowledges support from the Bosch-Forschungsstiftung. D.C. acknowledges the support of the Clore Scholars Programme and the Clore Israel Foundation. 
This work was supported by the European Union’s Horizon 2020 Research and Innovation Program under Grant Agreement No. 820394 (ASTERIQS) and QuantERA II under Grant Agreement No. 101017733, DFG (CRC 1279 and Excellence cluster POLiS), ERC Synergy Grant HyperQ (Grant No. 856432), BMBF and VW Stiftung. A.R. acknowledges the support of ERC Grant QRES, Project No. 770929, Grant Agreement No. 667192 (Hyperdiamond), ISF and the Schwartzmann university chair. F.J. acknowledges the support from the European Union projects QuMicro (Grant No. 101046911), QCIRCLE (Grant No. 101059999), C-QuENS (Grant No. 10113535) and the Carl Zeiss Stiftung.
We thank Dan Gluck for a useful discussion regarding our statistical analysis. 
\end{acknowledgments}


%
\newpage

\appendix

\onecolumngrid

\section{Experimental setup}\label{AppendixA}

Experiments for the correlation spectroscopy (CS) and Qdyne measurements have been performed on two different but conceptually equivalent setups. The shallow NV centers are addressed via a fluorescent confocal scan done on a home-built confocal microscope. The NV centers are initialized and read out using a $532\,\mathrm{nm}$ (CS) and $517\,\mathrm{nm}$ (Qdyne) laser pulse of about 5000\,ns (CS) and 1000\,ns (Qdyne) duration generated by a CW laser (Laser Quantum Gem532, Toptica iBeam smart 515) and chopped by an acousto-optic modulator (CS, Crystal Technology 3200-146) or by the pulsed-laser output itself (Qdyne). 
The pulse sequences containing both the trigger signals for laser output pulses, and microwave waveforms, are sampled on an AWG (Tektronix AWG70000A, Keysight M8195A), amplified (Amplifier Research 60S1G4, Amplifier Research 30S1G6) and applied to the NV center through a copper wire of $20\,\mathrm{\mu m}$ diameter strapped across the diamond sample. A single photon counting module (SPCM, Excelitas SPCM-AQRH-4X-TR) is used for detecting the photoluminescence (PL). The SPCM  generates a TTL output whenever it detects a photon.  These signals are recorded with a multiple-event-time digitizer (FAST ComTec P7887, FAST ComTec MCS6A) with time stamp with a timing resolution of 200\,ps. We use the Qudi software suite to orchestrate and control the experiment hardware \cite{Qudi}. 

To produce the diamond samples used in the experiments, natural abundance \textsuperscript{13}C diamond substrates were used on which an isotopically purified (99.999\,\% \textsuperscript{12}C) layer of about 150\,nm thickness was grown in a home-built plasma enhanced chemical vapour deposition growth reactor \cite{osterkamp2019engineering,findler2020indirect}. Shallow NV centers were subsequently created by ion implantation of \textsuperscript{15}N\textsuperscript{+} at a dose of $5 \times 10^{8}\,\mathrm{N^+\,cm^{-2}}$ and implantation energies between 2 and 5\,keV. Slightly deeper NV centers were created using the indirect overgrowth method as described in \cite{findler2020indirect} using an ion dose of $1 \times 10^{11}\,\mathrm{N^+\,cm^{-2}}$ and 2.5\,keV implantation energy. The diamond samples were then annealed in a home-built UHV furnace at $1000\,^\circ\mathrm{C}$ for 3 hours to provide the combination of stable nitrogen-vacancy pairs by mobilizing vacancies and, in addition, to heal radiation damage \cite{lang2020long}. The samples were regularly cleaned for 30 minutes in a equal mixture of sulphuric (97$\%$), perchloric (70$ \%$) and nitric acid (65$\%$) at a process temperature of $200 \,^\circ\mathrm{C}$ in a microwave reactor system (MWT AG, type ETHOS Lab). 

The NV  center depths were estimated by measuring the power spectrum of the hydrogen nuclei in the immersion oil with dynamical decoupling sequences \cite{Viola1999,Cywinski2007,KDDref,KDDref2,CasanovaPRA2015,Suter2016,GenovPRL2017}. The power spectrum measurement allows estimation of $B_\mathrm{rms}$ produced by the nuclear spins at the surface of the diamond. With this one can estimate the depth $d_\mathrm{nv}$ of the NV center from the formula \cite{Pham2016}: 
\begin{align}\label{Eq:phi_rms_res}
B_{\text{rms}}^2=\rho\left(\frac{\mu_0\hbar\gamma_n}{4\pi}\right)^2\left(\frac{5\pi}{96d_\mathrm{nv}^3}\right),
\end{align}
where $\rho$ is the nuclear spin number density, $\gamma_n \approx 2.68\times10^8\,\mathrm{rad\,s^{-1}T^{-1}}$ is the nuclear spin gyromagnetic ratio for hydrogen, and the term in the second brackets is a geometric factor which assumes that the NV center is oriented at an angle of 54.7\,$^{\circ}$ with respect to the normal to the surface, so the latter is along the [100] crystal direction.

\section{Frequency undersampling}\label{AppendixB}

In both the CS and Qdyne experiments we choose particular sampling times, so the signal is undersampled and the detected frequency is much smaller than the actual Larmor frequency. Given a certain signal oscillating at a (Larmor) frequency $f_\mathrm{L}=\omega_\mathrm{L}/2\pi$ (in frequency units), and a target undersampling frequency $f_\delta=\delta/2\pi$ (in frequency units), where $f_\delta\ll f_\mathrm{L}$, the minimum undersampling time step is given by 
\begin{equation} \label{t_undersampling_min}
    t_{s,\mathrm{min}}= \frac{1}{f_\mathrm{L}-f_\delta}.
\end{equation}
The actual sampled times depend on the number of points we want to sample within a single period of the undersampling frequency. 
For example, in our experiments we typically have $n_{s}\sim 10$ sampled points per cycle of the undersampling frequency ($n_{s}= 2$ is a minimum) to probe the signal, so the actual sampling time step is
\begin{equation} \label{t_undersampling}
    t_{s}= \underbrace{\round{\frac{1}{n_{s}-1}\frac{t_{\delta}}{t_{s,\mathrm{min}}}}}_{k} t_{s,\mathrm{min}},
\end{equation}
where $t_{\delta}=1/f_\delta$ is the period of the undersampling frequency and $k\ge 1$ is an integer, which rounds the expression $\frac{1}{n_{s}-1}\frac{t_{\delta}}{t_{s,\mathrm{min}}}$ to obtain approximately $n_s$ sampled points within $t_\delta$.

\section{Correlation spectroscopy}\label{Appendix:CS_single_NV}

\subsection{Description of the protocol}

The correlation spectroscopy measurements were carried out on single NV centers with depths ranging from $d_\mathrm{nv} \approx 3$\,nm to $\approx 14$\,nm below the diamond surface. 
We used the Knill dynamical decoupling (KDD4) pulses sequence for our experiments \cite{KDDref,KDDref2,CasanovaPRA2015,GenovPRL2017}. Each KDD has five $\pi$ pulses with phases 
$(\alpha+\pi/6,\alpha,\alpha+\pi/2,\alpha,\alpha+\pi/6)$ 
We obtain the KDD4 by nesting a KDD within the XY4 sequence, 
meaning that the phase $\alpha$ takes values $(0,\pi/2,0,\pi/2)$, resulting in a sequence of twenty pulses. We repeat it $N$ times, which we label as KDD4-N 
where N is the order. 
The pulses sequence is
\begin{gather}
\pi/2(x)- \text{DD sequence for time $\tau$} -  \pi/2(y) \notag\\
- \text{waiting time $\tau_\mathrm{w}$} -  \notag\\
\pi/2(x)- \text{DD sequence for time $\tau$} -  \pi/2(\pm y),
\end{gather}
where the two versions of the last pulse $\pi/2(\pm y)$ give the alternating projections on the $|0\rangle$ and $|1\rangle$ states, respectively. 
%
%
The measured signal is the difference between the two alternating measurements and takes the form
\begin{align}
c_{\text{cs}} &= c_\mathrm{max} \sin\left(\Phi_1\right)\sin\left(\Phi_2\right)\approx c_\mathrm{max} (\Phi_1 \Phi_2), 
\end{align}
where $c_\mathrm{max}$ is the maximum contrast, $\Phi_1$ ($\Phi_2$) is the accumulated phase during the first (second) DD sequence, and the last approximation, $\sin\left(\Phi_j\right)\approx \Phi_j,j=1,2$, is valid for small phases and is fulfilled in our experiment. We perform multiple readouts and average the result, which leads to
\begin{align}
\overline{c}_{\text{cs}}=c_\mathrm{max} \langle\Phi_1 \Phi_2\rangle= c_\mathrm{max}  \frms^2 \cos{(\omega t)}C(t),
\end{align}
where $t=\tau+\tau_\mathrm{w}$, $C(t)$ is the envelope of the autocorrelation and $\frms^2$ is the phase variance. An exponential decay envelope is  often assumed in NV NMR but it has been recently shown that the envelope follows a power-law decay \cite{Cohen2020,Staudenmaier2022}. More details on the experimental procedure with correlation spectroscopy can be found in Ref.~\cite{Staudenmaier2022}.

\subsection{Signal-to-noise ratio}

The signal-to-noise ratio (SNR) is primarily determined by photon shot noise, and depends on the photon count rate and the number of measurements \cite{Schmitt2016diss}
\begin{equation}\label{eq:SNR}
    SNR =\frac{N\left(\eta_0 - \eta_1\right)}{\sqrt{N\left(\eta_0+\eta_1\right)}}=\sqrt{N}\frac{\eta_0 - \eta_1}{\sqrt{\eta_0+\eta_1}},
\end{equation}
where $\eta_0$ ($\eta_1)$ is the expected photon count per measurement when NV is prepared in the $m_s=0$ ($m_s=\pm1)$ state, and $N$ is the number of measurements. It is evident that the $SNR\propto\sqrt{N}$, so it is beneficial to perform more measurements per data point. 
The spin dependent relative fluorescence contrast is $\chi=\frac{\eta_0-\eta_1}{\eta_0}$, and inserting it into Eq.~\eqref{eq:SNR} results in
\begin{equation} \label{snr2}
    SNR= \sqrt{N}\frac{\chi\sqrt{\eta_0}}{\sqrt{2-\chi}}.
\end{equation}

As per Eq.~\eqref{snr2}, the SNR can be improved by increasing the expected number of photons per measurement $\eta_0$, the relative contrast $\chi$, and the number of measurements $N$. One can increase $\eta_0$ by prolonging the readout duration but, at some point, the contrast $\chi$ saturates and then starts decreasing, so an optimal readout duration depends on the specific experiment and is typically in the range of a few hundred nanoseconds. In a standard single NV experiment, one can obtain a maximum contrast of about $30\,\%$ where the count rate is typically $\eta_0\approx 0.03 - 0.1$ photons per measurement. 

An alternative way to improve the SNR is to repeat each measurement many times, i.e., increase $N$. However, this leads to an increase in the acquisition time per data point. In order to have a moderate SNR within a reasonable total measurement time we carefully choose the number of sampled times 
by undersampling. This allows to reduce the number of sampled points to match a certain number of samples within the period of a chosen undersampling frequency $f_\delta$. Thus, within a fixed total measurement time the number of measurements per sampled time is increased to have a good SNR.

\section{Qdyne}\label{Appendix:Qdyne}
\subsection{Experimental requirements for Qdyne}

First, we emphasize that Qdyne measurements require the same hardware as ``conventional'' pulsed quantum sensing experiments, e.g., power spectrum measurements or correlation spectroscopy. Any pulse generator or arbitrary waveform generator  can be used for these measurements; the only requirement is to have a precise rate with which the single measurement sequences are executed. The second essential device is a (fast) counting card. This has to register the arrival time of photons, in case of a digital detector signal, or associate an analog signal with time. The timing resolution has to be such as to resolve the readout laser pulse properly -- a property necessary for standard pulsed measurements as well. 

The Qdyne measurement sequence in itself does not differ much from the conventional type of measurements. However, there are major differences in data handling and analysis. The Qdyne sequence has no sweep parameter (as opposed to CS where the wait time $\tau_\mathrm{w}$ is swept). Instead, a single (fixed) measurement is repeated many times. Importantly, the number of photon counts of every single measurement are stored individually in a long time trace. By the constant measurement rate temporal correlations are imprinted onto this readout data. 
These correlations can be substantiated during data postprocessing using autocorrelation processing and/or Fourier transform analysis, for example. 

Basically, this is true for any kind of Qdyne measurement independent of the underlying signal \cite{Degen2017,Schmitt2017,Meinel2021,Staudenmaier2021, Staudenmaier2022}. While for coherent signals direct Fourier transform is suitable to obtain the signal, this is not true for signals with a short coherence (correlation) time, as we investigate in this work. 

\subsection{Optimization of the nanoscale NMR Qdyne measurements}\label{Qdyne_optim}
In the following we describe in detail our procedure for optimization of the Qdyne measurements in the setting of liquid, statistically polarized nanoscale NMR. \\

\textbf{Dynamical decoupling time} \\

Recent theoretical work shows that the optimal 
Qdyne measurement duration requires $\Phi_\mathrm{rms} \approx 1$ for small undersampling angular frequency $\delta$, such that $(\delta/2\pi)\,T_D < 1$ \cite{Oviedo2020}. Specifically, decoherence of the NV sensor and the limited coherence time of the signal have to be taken into account when choosing the optimal dynamical decoupling (DD) duration and thus 
$\Phi_\mathrm{rms}$. DD is usually designed such that an accumulated phase is $\Phi \propto B\tau$, where $B$ is the magnetic signal envelope during the DD sensing time $\tau$. In Qdyne, the phase is mapped onto a population difference, which scales as $\sim\sin(\Phi)$. At first, it might seem intuitive to perform many very short measurements to sample the signal better. However, a larger phase $\Phi$ would in principle increase the signal, especially when $\Phi \ll 1$. In addition, readout and initialization of the qubit sensor come with a temporal overhead. For example, the optical readout of the NV center is usually of the order of 3\,$\mu$s. Thus, increasing the number of measurements at the cost of a poor signal is typically not an optimal solution. On the other hand, prolonging the sensing time might lead to decoherence of the sensor qubit or lower signal due to diffusion. Hence a sweet spot has to be found that maximizes the acquired information by increasing the extractable signal and having a fast measurement rate. 

\textit{Derivation of the optimal DD duration} --- 
During each experimental run, the sensor accumulates a phase, which depends on the magnitude and phase of the nuclear spin signal, $\Phi_t=\frac{2}{\pi} \gamma_e B(t) \tau\cos{\left(\omega_\mathrm{L} t+\beta\right)}$, where the prefactor $2/\pi$ is due to the application of DD sequences, $\gamma_e$ is the gyromagnetic ratio of the NV center, $t$ is the starting time of the experimental run, $B$ is the signal amplitude, and $\beta$ is the signal phase at $t=0$. 
The probability that the sensor qubit is in state $|0\rangle$ ($|-1\rangle$) after the experimental run is $P_t=\frac{1}{2}-\frac{\sin{\Phi_t}}{2}$ ($P_t=\frac{1}{2}+\frac{\sin{\Phi_t}}{2}$). Then, if the system is initialized in state $|0\rangle$ the expected number of detected photons after the experimental run is given by 
\begin{equation}
\langle\eta_t\rangle \approx \eta-\frac{\widetilde{c}}{2}\sin(\Phi_t),\qquad\widetilde{c}=e^{-(\tau/T_2)} \, c
\end{equation}
where $\eta=(\eta_0+\eta_1)/2$ and $c=\eta_0-\eta_1$, $\tau$ is the duration of one DD sequence, and $T_2$ characterizes the decay of our signal due to decoherence of the sensor qubit during the DD sequence. The actual number of detected photons $\eta_t$ and the time $t$ are recorded for postprocessing. 

In order to analyze the data we calculate the covariance
\begin{equation}
\mathrm{Cov}\!\left( \eta_{t_0} \eta_{t_0+t} \right) = \left\langle \eta_{t_0} \eta_{t_0+t} \right\rangle -\left\langle \eta_{t_0} \right\rangle\, \left\langle \eta_{t_0+t} \right\rangle =
\frac{\widetilde{c}^2}{4} \left\langle \sin(\Phi_{t_0}) \sin(\Phi_{t_0+t}) \right\rangle \approx \frac{\widetilde{c}^2}{4} \left\langle \sin^2{\Phi_{t_0}} \right\rangle C(t/T_D)\cos{\left(\omega_\mathrm{L} t\right)},
\end{equation}
where we used in the last expression that $\Phi_t$ are typically small and that the signal in our experiments  is due to statistical polarization $\Phi_t\sim N(0,\Phi_\mathrm{rms})$ with $\Phi_\mathrm{rms} = \frac{2}{\pi} \gamma_e B_\mathrm{rms} \tau$ its standard deviation. In addition, $\left\langle B(t)B(0)\right\rangle\propto=B_\mathrm{rms}^2C(t/T_D)$ and $C(t/T_D)$ characterizes the decay of the signal autocorrelation due to diffusion with characteristic time $T_D$  \cite{Cohen2020,Cohen2020b,Oviedo2020,Staudenmaier2022}. The probability density function $\textrm{PDF}(\Phi_t; \Phi_\mathrm{rms}) = \frac{1}{\sqrt{2\pi} \frms} e^{-\Phi_t^2/2 \frms^2}$ centered at zero. 
One can estimate the variance  
\begin{equation}
    \left\langle\sin^2{\Phi_t}\right\rangle=\int_{-\infty}^{\infty} \sin^2(\Phi_t)\,\textrm{PDF}(\Phi_t; \Phi_\mathrm{rms}) \, d\Phi_t=\frac{1}{2}\left(1-e^{-2 \frms^2}\right)\approx \Phi_\mathrm{rms}^2,
\end{equation}
where the latter approximation is valid only for small $\frms$. The minimum time $t$ for calculating an autocorrelation function is the duration of a single measurement, i.e., $t\ge\tau$. Thus, in order to maximize our signal by choosing the appropriate duration of the DD sequence $\tau$, one has to maximize 
\begin{equation}
   \frac{\widetilde{c}^2}{8} C(\tau/T_D)\left(1-e^{-2 \frms^2}\right)=\frac{c}{8}e^{-(2\tau/T_2)} C(\tau/T_D) \left(1-e^{-2 \left(\frac{2}{\pi}\gamma_e B_{\text{rms}}\tau\right)^2}\right).
\label{Eq:Qdyne_signal}
\end{equation}
To get a qualitative measure of the SNR one has to multiply the signal of Eq.~\eqref{Eq:Qdyne_signal} with the square-root of the measurement rate $1/\sqrt{\tau_\mathrm{o} + \tau}$ with overhead time $\tau_\mathrm{o}$. 

\begin{figure}
    \centering
    \includegraphics[width=.9\linewidth]{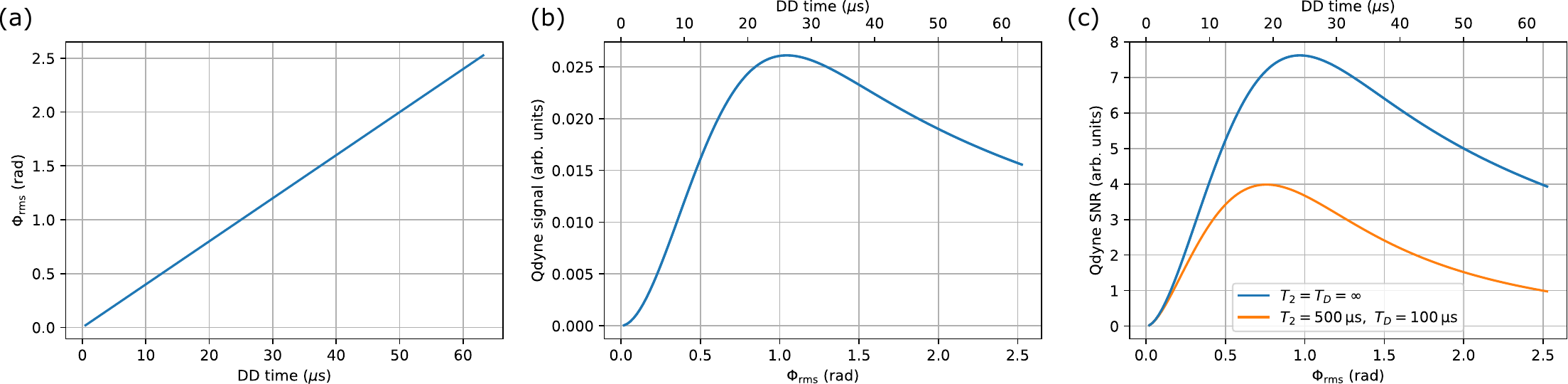}
    \caption{DD order optimization. Here a typical NV center at a depth of $d=8\,\mathrm{\mu s}$ with coherence time $T_2 = 500\,\mathrm{\mu s}$ is used. (a) $\Phi_\mathrm{rms}$ as function of the dynamical decoupling time. (b) The measurement contrast signal as of equation \eqref{Eq:Qdyne_signal}. (c) Eq. \eqref{Eq:Qdyne_signal} multiplied with the square-root of the sample rate $1/\sqrt{\tau_\mathrm{o} + \tau}$ with overhead time $\tau_\mathrm{o} = 3.5\,\mathrm{\mu s}$ to get a measure for the signal-to-noise ratio. The blue curve is calculated disregarding the coherence time of the NV center and the diffusion of the nuclear spins while the orange curve takes them into account. 
    The orange curve has its maximum at a little lower $\Phi_\mathrm{rms}$ indicating that a shorter dynamical decoupling time should be used.}
    \label{fig:DDoptimization}
\end{figure}

In Figure~\ref{fig:DDoptimization} we show the calculation that is done in order to find the optimal dynamical decoupling time for the Qdyne measurements using a given NV center. Specifically, in this example the NV center is at depth $d_\mathrm{nv}=8\,\mathrm{nm}$ and has a coherence time $T_2 = 500\,\mathrm{\mu s}$. The resulting diffusion time of the nuclear spins in the immersion oil is estimated to be $T_D = 100\,\mathrm{\mu s}$. Figure~\ref{fig:DDoptimization}(a) shows the increase of $\Phi_\mathrm{rms}$ with $\tau$. In part (b) and (c) the Qdyne signal as of Eq.~\eqref{Eq:Qdyne_signal} and the SNR are calculated, respectively. We find that the condition $\Phi_\mathrm{rms} \approx 1$ is matched when the finite coherence time of the NV center and the diffusion are disregarded. If they are considered, the best dynamical decoupling is found for shorter times. \\

We would like to note that this calculation here is akin to the theoretical calculations presented throughout the manuscript, done in order to maximize the Fisher information for the frequency. There, we consider a multi-parameter estimation problem, and calculate the Fisher information matrix. The diagonal entries of the inverse of the Fisher information matrix correspond to the different mean squared errors of the parameters to be estimated. Then, it can be shown that, in order to minimize the error in estimating the frequency, a $\frms \leq 1$ is optimal. Further details can also be found in \cite{Oviedo2020}. \\

\newpage
\textbf{Magnetic field stability} \\

The magnetic field has to be stable throughout the whole duration of the experiment. As the measurement is quite time consumptive this might be a limiting factor to the final spectral resolution of the signal. The magnetic field might fluctuate or drift due to mechanical vibrations in the setup or because of temperature shifts that affect both the thermal expansion of the mechanical mountings and the magnetization of a permanent magnet. The gyromagnetic ratio of the proton is $\gamma_{^1\!H} = 42.6\,\mathrm{MHz\,T^{-1}}$, hence a shift of the magnetic field by 0.1\,G results in a shift of the Larmor frequency of 426\,Hz. For that reason the magnetic field has to be kept stable with sub-Gauss precision. Having a high stability magnet that decouples from environmental conditions might be beneficial. 

Here we use a permanent magnet made of either neodymium or samarium cobalt. The latter one has a lower temperature coefficient of remanence, –0.09 to –0.12\,\%/K for neodymium and -0.03 to –0.05\,\%/K for samarium-cobalt magnets. During the Qdyne measurements the magnetic field is monitored by performing pulsed ODMR for both $m_S = \pm1$ transitions. From both resonances the strength of the magnetic field is determined and shifts during the measurement are recorded. With that we can slightly adjust the sample rate of the AWG in order to keep the ratio between the measurement rate of the Qdyne measurements and the Larmor precession constant. Assuming a small drift in the magnetic field results in a shift of the initial Larmor frequency $f_\mathrm{L}$ of $\Delta f_\mathrm{L}$. We can then change the sample rate $f_\mathrm{samp}$ by an amount $\Delta f_\mathrm{samp}$ such that the ratios $\frac{\Delta f\mathrm{samp}}{f_\mathrm{samp}} = \frac{\Delta f_\mathrm{L}}{f_\mathrm{L}}$. As the shift is generally very small ($<$ 0.1\,\%) this won't affect the measurement any further. 

\begin{figure}
    \centering
    \includegraphics[width=.9\linewidth]{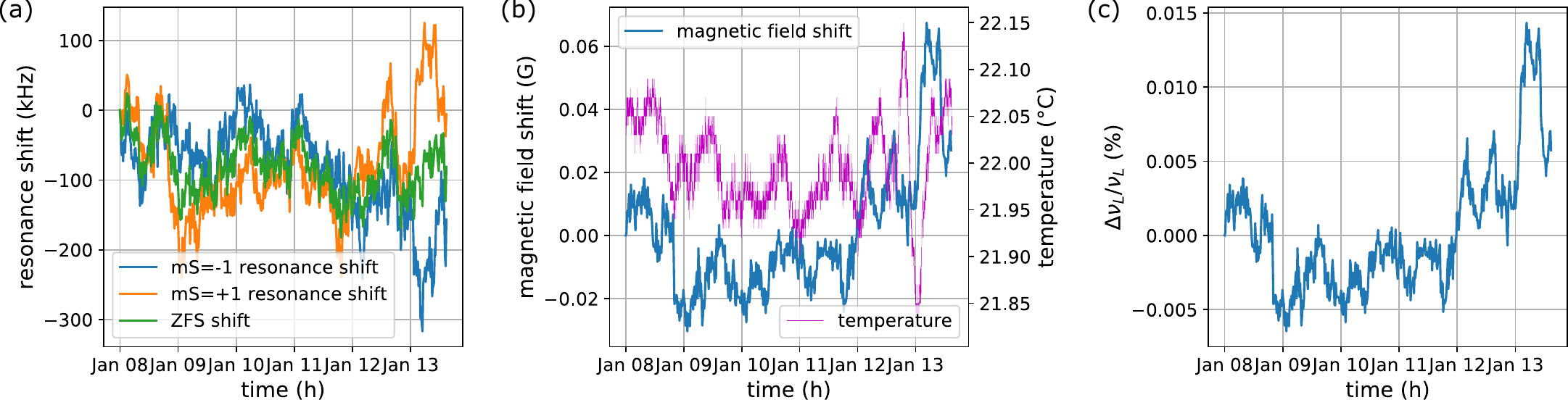}
    \caption{Magnetic field monitoring. (a) Shift of the $m_S=-1$ (blue) and $m_S=+1$ (orange) resonance frequencies obtained from pulsed ODMR measurements and the corresponding shift of the ZFS. (b) Calculated absolute drift of the magnetic field at 480\,G and the recorded temperature. No clear correlations between the two are discovered. (c) Relative shift of the Larmor precession frequency. The sample rate of the AWG can be adjusted by the same relative amount to compensate for the change in the frequency.}
    \label{fig:magnet_monitoring}
\end{figure}

In Figure \ref{fig:magnet_monitoring}a we see the results of the pulsed ODMR measurements throughout the whole set of Qdyne measurements. After 15 minutes of Qdyne measurements the magnetic field has been checked with low mw power pulsed ODMR. One can see that the drift of the resonance frequency is mostly due to shifts of the zero-field splitting (ZFS). The remaining shift can be attributed to a change in the strength of the magnetic field (Fig. \ref{fig:magnet_monitoring}b). Here, the applied bias magnetic field is 480\,G where the NV center resonance frequencies are 1531\,MHz and 4209\,MHz for the $m_S = -1$ and $+1$ transition, respectively. The temperature at the magnet is recorded as well. However, we cannot find clear correlations between the temperature and a shift in the magnetic field. As described above we can compute the expected shift in the Larmor frequency and from that determine the change in the sample rate of the AWG. With our AWG (Keysight M8195A) we use a the sample rate of 64\,GS/s that can be adjusted with 10\,mHz precision. Note that we assume that the direction of the magnetic field stays constant as usually the field strength is more subject to fluctuation than the direction. Furthermore, the observed shift of the ZFS cannot be explained by an asymmetry of the resonances due to a change of the direction and misalignment of the field. \\

\textbf{Readout duration} \\

Determining the best readout parameters for the measurement and data analysis is crucial. We see from the Fisher information [Eq.~(4) in the main text or Eq.~\eqref{eq:FI_Qdyne} below] that Qdyne measurements suffer even more from reduced contrast of the readout and from overhead time. So first of all, the latter should be reduced. For that the laser pulse for optical pumping (initialization) of the NV center into the $m_S = 0$ spin state should be not longer than necessary and wait times just as long as required. Second, we try to optimize the readout by precise characterization of the collection window of photons during the laser pulse. While the laser pulse is long enough in order to guarantee initialization of the NV center in the $m_S=0$ state, fluorescence dependent information of the spin state is only acquired during a shorter time of the laser irradiation. 

\begin{figure}
    \centering
    \includegraphics[width=.9\linewidth]{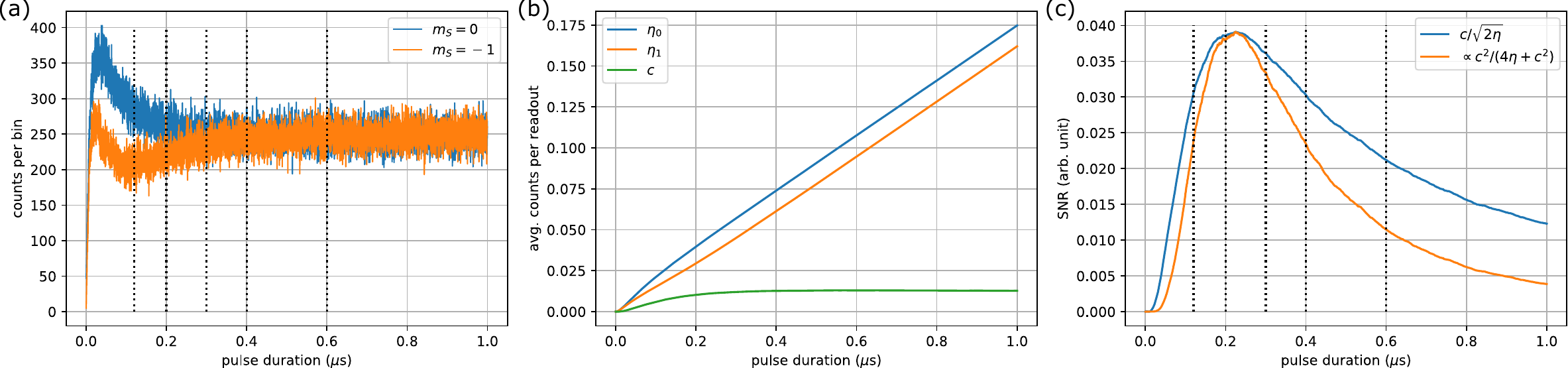}
    \caption{Readout photon collection window optimization. (a) Fluorescence curve of the NV center $m_S = 0$ (blue) and $m_S = -1$ (orange) spin state in dependence on the laser pulse duration. (b) Average counts per readout (orange and blue) and readout contrast (green). (c) Readout quality determined by the SNR of a single readout as of Eq.~\eqref{eq:SNR} (blue) and by $\frac{c^2}{4\eta + c^2}$ (orange) as the scaling of the Fisher information. Both calculations find their maximum at the same position. The latter is re-scaled to match the maximum values. The dashed lines represent the stop of the photon collection window for the white data points shown in Figure~2(b) in the main text.}
    \label{fig:readout_optimization}
\end{figure}

We perform separate readout of the NV center $m_S = 0$ and $m_S = -1$ spin state and record the fluorescence as a function of the laser pulse duration as shown in Figure~\ref{fig:readout_optimization}(a). The cumulative sum of the photon counts divided by the number of performed readouts gives the average photon counts $\eta_0$ and $\eta_1$ in dependence of the readout duration (b). The difference and the average of both give the contrast $c$ and average photon count rate $\eta$ which determine the quality of the readout. Above in Eq.~\eqref{eq:SNR} it is argued that $SNR = \frac{c}{\sqrt{2\eta}}$ for the single laser readout while in the main text we outline that the Fisher information scales with the readout dependent factor $\frac{c^2}{4\eta + c^2}$. In Figure~\ref{fig:readout_optimization}(c) we plot both values and find that both maximize at the same laser pulse duration, even though showing slightly different curves. By maximizing either of the two values the best photon collection window for the readout is determined. The Qdyne data time trace is now set up by counting the number of photons that arrive during each readout within the determined collection window. As the average count rate $\eta\sim0.04$ it is mostly zeros, however higher counts such as two and three photons are possible with a low probability. For the data presented as white shapes in Figure~2(b) in the main text the collection window is varied. In Figure~\ref{fig:readout_optimization} panel (a) and (c) this is indicated by the vertical dashed lines. 

\subsection{Qdyne autocorrelation}
\begin{figure}
    \centering
    \includegraphics[width=.9\linewidth]{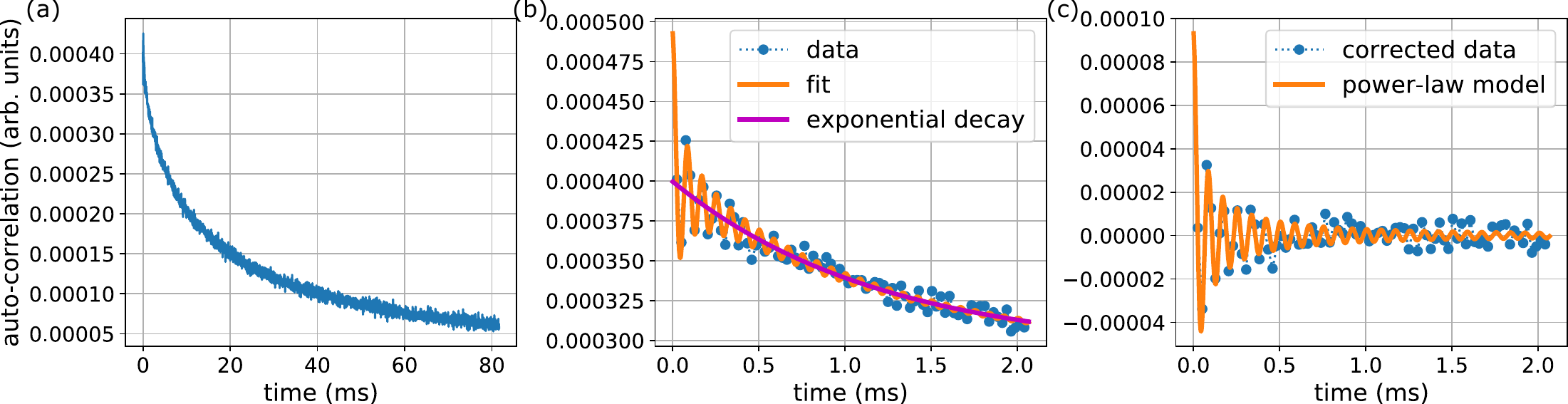}
    \caption{Auto-correlation of a Qdyne measurement. (a) Auto-correlation of the time trace recorded from a Qdyne measurement. (b) Auto-correlation with a fit including the signal model function plus an additional exponential decay. (c) Corrected autocorrelation data where the exponential decay is removed and a fit with only the signal model function is performed.}
    \label{fig:qdyne_ac}
\end{figure}

Above it is described how the time trace for the Qdyne measurement is obtained. For the analysis of our data we calculate the autocorrelation of the time trace using the \textsl{statstool} package in Python. The autocorrelation then shows several decays. These result from different potential noise sources. First of all, the target signal resulting from the Larmor oscillation decays as described in the main text. Additionally, noise from the readout is captured as a decay in the autocorrelation as well. This can be laser power fluctuations, fluorescing particles in the immersion oil that diffuse through the laser beam and conversion of the charge state of the NV center. The first one fluctuates on a time scale of seconds, for what reason it is not further important here. However, a decay of the autocorrelation on the order of a few tens of milliseconds can be found [Figure~\ref{fig:qdyne_ac}(a)]. We attribute this decay to single fluorescing particles in the immersion oil. 

In order to remove the effect of the additional decay we perform a fit with a model which includes a term accounting for exponential decay in addition to the signal term, $\Phi_\mathrm{rms} \cos(\delta t) C(t/T_D) + A \exp(-t/T_\mathrm{exp}) + \mathrm{offset}$. This is shown in panel (b) of Figure~\ref{fig:qdyne_ac}. The offset accounts for other decay of the autocorrelation on a long time scale which is not of relevance considered here. In (c) the autocorrelation data is shown where the exponential decay is removed and a fit with the pure model function, $\Phi_\mathrm{rms} \cos(\delta t) C(t/T_D)$, is performed. The measurement presented here corresponds to the data present also in Figure~2(a) in the main text. 

\section{Theoretical considerations}\label{AppendixE}

In this section, we thoroughly explain several technical details regarding the usage of the Fisher information as an optimization tool, clarify the theoretical points behing the usage of the different data analysis algorithms,  and provide a mathematical derivation of the Fisher information for the power-law autocorrelation.

\subsection{Assumptions and limitations concerning the Cram\'er-Rao bound}\label{CR}

In the following we would like to address some technical issues and assumptions concerning the use of the Cram\'er-Rao bound and the Fisher Information, as presented in the main text.

First and most importantly, the main assumption of the Cram\'er-Rao bound as an inequality in this form, e.g. $\Delta\delta\leq 1/I_\delta$, where $\Delta\delta$ is the variance (or MSE) of the frequency estimator for $\delta$ and $I_\delta$ is the FI about $\delta$, is that the estimator is unbiased. Therefore, the Cram\'er-Rao bound provides a lower bound on the MSE as long as our measurement scheme and analysis do not suffer from an intrinsic bias that would systematically shift the spectrum. The measurements are performed using dynamical decoupling sequences, which remove most of the static noise components that might shift the spectrum and the remaining drift is minimized by periodic recalibration, an important feature for successful nano-NMR Qdyne experiments, as we explain in Appendix \ref{Qdyne_optim}. The analysis is performed by using nonlinear least squares. Arguably, this could create large errors due to multiple local minima. This is avoided both by repeating the procedure multiple times to guarantee convergence and by using the binning analysis of different subsets of the data for the final frequency estimation.

We do have to concede that the Cram\'er-Rao bound derived in this work might not be achievable as an equality, because it does not take into account all possible noise sources that are included in the experiment. However, trying to account for all possible noise sources, in order to saturate the corresponding bound, can quickly turn out to be quite impractical and of little value for the following reasons:

- In our derivation of the Fisher information, we take into account the main noise source for the sample in the experiment, which is diffusion, and it is accounted for exactly, and the main noise source for the probe, which is photon shot noise, which is also accounted for exactly.  Dynamical decoupling makes all other error sources small in comparison, and is also exactly considered in the derivation. For these two assumptions, the Fisher information theoretically derived in this work captures the main features of the experiment and is, therefore, a good measure with which to benchmark the experimental MSE. 

- Moreover, the experimental setups for both protocols are quite similar and are, therefore, prone to feature similar errors not accounted for by the theory. These errors would consequently produce a deviation from the possible saturation of the theoretically calculated Cram\'er-Rao bound, as is apparent in Fig. 2 of the main text, and which accounts for the discrepancy observed. 


- Finally, we would like to mention that it is possible, in principle, to account for other error sources in the FI calculation by assuming that noisy experimental parameters are also parameters that require an estimation, thereby adding some more information to the estimation of the frequency, however, as we mentioned, this turns out to be of little value. Trying to do it means using multi-parameter estimation, for which the Cram\'er-Rao bound becomes a matrix inequality for the covariance matrix, and the noise can be inserted as the standard deviation of the estimated parameters. This, however, becomes quickly impractical, because it requires inverting the large covariance matrix (consisting of all possible noise channels), while we gain only limited additional information.  

To sum up, we do not claim to be able to saturate the theoretical Cram\'er-Rao bound. However, we claim that we can use it in a mathematically consistent procedure with which to optimize experimental scenarios, measure for the best possible performance of a given protocol, and compare protocol performances, as we demonstrate.

\subsection{Data analysis algorithms}\label{MLE}

Under the assumption that the noise in each measurement is Gaussian with zero mean and a constant variance, MLE and least squares are mathematically equivalent, as we will show in the following.

For a given set of measurements $\left\{x_i,y_i\right\}_{i=1}^N$ and a fit of the form $\bar{y}=f\left(\bar{\theta},\bar{x}\right)$, where $\bar\theta$ are the fitting parameters, the least squares estimator is given by

\be
\bar{\theta}_{LSQ}= \text{arg min}_{\bar\theta} \sum_{i=1}^N \left[y_i-f(\bar\theta,x_i)\right]^2.
\ee

Assuming the noise in each measurement is Gaussian with zero mean and a constant variance we have

\be
P(\bar{y}|\bar{\theta},\bar{x}) = \prod_{i=1}^N\frac{1}{\sqrt{2\pi\sigma^2}}\exp\left(-\frac{\left(y_i-f(\bar{\theta},x_i)\right)^2}{2\sigma^2}\right).
\ee

Therefore, the MLE is given by

\be
\bar{\theta}_{MLE}=\text{arg max}_{\bar{\theta}} \log P(\bar{y}|\bar{\theta},\bar{x}) = \text{arg min}_{\bar{\theta}} \sum_{i=1}^N \left(y_i-f(\bar{\theta},x_i)\right)^2 = \bar{\theta}_{LSQ},
\ee
where in the second equality the change between arg min and arg max is justified by the minus sign of the Gaussian.

This simple derivation can be generalized using weighted least squares for any distribution within the exponential family
\cite{Charnes76}.

Even the simple Gaussian noise is a reasonable modeling for our system, since we eliminate the bias by using dynamical decoupling and by periodically calibrating the system and other noise sources are of constant amplitude throughout the measurement (e.g. amplitude noise, remaining  magnetic fluctuations, timing errors etc.). In that sense, there is no additional error introduced by using nonlinear least squares instead of MLE. 

In this case, the technique of choice depends more on the characteristics of the signal that one wants to analyze, and the algorithm's convergence rate for each of the possible scenarios. For a purely coherent signal, the scaling of the Fisher information is $T^3$
for either estimating the frequency from the correlation function that includes all possible correlated points, or from analyzing the measurement vector directly, which is the outcome of a Qdyne experiment and, therefore, the natural result to analyze. The former case calls for the use of least squares analysis. The second, since it involves analyzing each measurement in terms of its outcome probability, is better done with MLE. Computationally, both behave similarly in terms of convergence rate, since they require minimizing similar functions.

For signals with limited coherence, the situation is different. There, the parameter space describing the measurement vector grows linearly with the total measurement time, as parameters change, roughly, every coherence time. Then, implementing an MLE analysis on the measurement vector quickly becomes computationally demanding, imposing severe restrictions over the precision that it is possible to achieve in a reasonable time. On the other hand, a least squares analysis performed on the autocorrelation of said measurement vector, remains computationally amenable and, therefore, can be forced to the desired numerical precision, provided sufficient data and using techniques that guarantee convergence, as we described in the previous comment. It is important to stress that, from a purely mathematical perspective, nothing precludes one from using MLE on the measurement vector and, in fact, it could lead to small gains whenever the signal-to-noise ratio is poor, as the autocorrelation would become quite noisy quite rapidly, masking the oscillation. However, even this small gain gets overshadowed in most signals by the computational time it takes to run an effective, precise MLE algorithm.  

Additionally, we would like to note that, while it is true that the nonlinearity of the least squares analysis makes it more prone to fall onto local minima, the same is true for MLE analysis, for which the probability function describing the experiment is highly nonconcave.

\subsection{Fisher information for power law correlations}\label{AppendixQdyneInfo}

Here, we detail the calculations that lead to the Fisher information equations, for correlation spectroscopy and Qdyne, in the case of the power law correlations, shown in the main text. The signal that is probed by the NV center originates in the statistical polarization of a small distribution of nuclei, and can be described as $B(t) = a(t)\cos(\delta t) + b(t)\sin(\delta t)$. The covariance of this signal is influenced by the fluctuation of the amplitudes $\{a(t),b(t)\}$ due to noise, such that ${\text{cov}}\langle\Phi[B(t_0)]\Phi[B(t_0+t)]\rangle = \Phi_{\text{rms}}^2\cos(\delta t)C(t/T_D)$, where $t$ is a variable time separation between measurements. For correlation spectroscopy, $t=\tau+ \tau_\mathrm{w}$ with $\tau$ the duration of a phase acquisition period (or dynamical decoupling time), and $\tau_\mathrm{w}$ a variable waiting time in between two phase acquisition periods. For Qdyne, $t = n(\tau + \tau_\mathrm{o})$, with $\tau$ the phase acquisition period, overhead time $\tau_\mathrm{o}$ and $n$ an integer number representing an arbitrary separation between any two measurements. $C(t/T_D)$ is an envelope whose shape depends on the specific process that describes the fluctuations of $\{a(t),b(t)\}$. For nanoscale liquid samples, fluctuations are mostly due to diffusion of the molecules in the sample, which can be described as a random process with zero mean and $T_D$ correlation time, where $T_D = d_\mathrm{nv}^2/D$. Then, 
\begin{equation}
C(z) = \frac{4}{\sqrt{\pi}} \Bigg(
 z^{-\frac{3}{2}} - \frac{3}{2}z^{-\frac{1}{2}} + \frac{\sqrt{\pi}}{4} + 3\sqrt{z} - \frac{3\sqrt\pi}{2}z + \\
 \sqrt{\frac{\pi}{z}}{\text{erfc}}\Big(z^{-\frac{1}{2}}\Big)\exp{z^{-1}} \times \\
 \Big( - z^{-\frac{3}{2}} + z^{-\frac{1}{2}} - \frac{7}{4}\sqrt{z} + \frac{3}{2}z^{+\frac{3}{2}} \Big)  \Bigg),
\label{Cohenian}
\end{equation}
with $z=t/T_D$ \cite{Cohen2020}.

The phase that the NV center accumulates depends both on $B(t)$ and on the response function $h(t)$ of the NV center to the series of pulses comprising the dynamical decoupling sequence of duration $\tau$, such that 
\be
\Phi[B(t')] = \gamma_n \int_{-\tau/2}^{\tau/2} ds\, h(s)B(t'+s).
\ee
If $\tau \ll T_D$, assuming pulses of negligible duration, the response function simplifies to $h(t') = \theta(t'+\tau/2)\theta(t'-\tau/2)$, and the covariance between two phases $\Phi$ taken at different times is ${\text{cov}}\langle\Phi[B(t')]\Phi[B(t'+t)]\rangle = \gamma_n^2\tau^2\sinc^2(\delta\tau/2){\text{cov}} \langle B(t')B(t'+t)\rangle \approx \frms^2\cos(\delta t)C(t/T_D)$, with $\frms = \frac{2}{\pi}\gamma_n\brms\tau$

Correlation spectroscopy combines two phase acquisition periods $\tau$ separated by a variable delay time $\tau_\mathrm{w}$, such that at the end of the second period, an interrogation of the state of the NV yields information about ${\text{cov}}\langle\Phi[B(t')]\Phi[B(t'+t)]\rangle$. At the interrogation time, the expected photon number is $p = \eta + c\langle \sin(\Phi_0)\sin(\Phi_t)\rangle/2$, where the coefficients $\eta$ and $c$ represent the average photon number and contrast, respectively, and are defined through the fluorescent rates of each of the NV states as $\eta = (\eta_0 + \eta_1)/2$ and $c = \eta_0 - \eta_1$. For a weak signal, the sine can be approximated by the angle and we have that $p \approx \eta + c \frms\cos(\delta t)C(t/T_D)/2$.

Given the probability of detecting a photon, we can calculate the Fisher information for the parameter of interest --the frequency $\delta$ in this case-- that such a measurement yields as
\begin{equation}
i_\delta = {\mathbb{E}}\left[\left(\frac{d\log(L(\delta))}{d\delta}\right)^2\right],\label{singleFIappendix}
\end{equation}
which for the probability defined above is \cite{Oviedo2020}
\be
i_{\delta} \approx \frac{c^2}{4 \eta +c^2}\frms^4 t^2 \sin ^2 (\delta t ) C^2( t/T_D ),
\ee
where we neglect higher order terms in $\frms$ by assuming that the amplitude of the signal is smaller than one.

We wish to obtain the total Fisher information gathered on an experiment featuring $N$ measurements, spanning a total experimental time $T$. The Fisher information is an additive quantity, so the total Fisher information, $I_\delta$, is the sum of the Fisher information of  each individual measurement $i_\delta$. Considering an average measurement time $T_D$ we can write, for each measurement $j$ starting at $t_j= j T_D$,  
\be
I_\delta^\mathrm{CS} = \frac{c^2}{4 \eta +c^2}\frms^4\sum_{j=0}^{T/T_D}  (jT_D)^2 \sin ^2 (\delta j T_D ) C^2( j).
\label{sumcs}
\ee
We can see that, for a $C(j)\sim j^{-3/2}$ such as for diffusion dominated noise, only the first terms in $j$ will contribute significantly to the total Fisher information, as the summand in Eq.~\ref{sumcs} scales with $j^{-1}$. This means that, since CS proves the autocorrelation directly, it is more advantageous to focus on the early time, with $\tau_\mathrm{w} \sim T_D$, such as it is explained above. However, this also shows that when $T_D$ is short compared to the frequency of the signal, it will not be possible to have a sufficient number of significant measurements, leading to the resolution problem. To see this explicitly we can calculate the sum: For a large number of measurements, namely in the limit $T \gg T_D$, we do the calculation transforming the sum to an integral form of a Riemann sum. To do so, we choose $j = T/T_D$. Taking the lower limit at $0$ fixes also the upper limit to $1$, and yields the following integral 
\be
I_\delta^\mathrm{CS} = \frac{c^2\frms^4 T_D T}{4 \eta +c^2}\int_{0}^{1} dt \,  t^2 \sin ^2 (\delta t T_D ) C^2( t ).
\ee
Solving analytically the above equation for $C(t)$ as in Eq.~(\ref{Cohenian}) is complicated. On the main text we do it numerically, but to understand the limits treated here, i.e. the low frequency limit $(\delta/2\pi) T_D < 1$ and the long experiment limit $(\delta/2\pi) T > 1$, we can approximate Eq.~(\ref{Cohenian}) by $C(t > T_D) \sim t^{-3/2}$, which describes the behavior of correlations at long times. Since most of the information is gathered from the long decay tail of the correlations, it describes the dominant term in the exact result. Then
\be
I_\delta^\mathrm{CS} \approx \frac{c^2\frms^4 T_D T}{4 \eta +c^2}\int_{0}^{1} dt \,  t^{-1} \sin ^2 (\delta t T_D ) = \frac{2}{\sqrt{\pi}}\frac{c^2\frms^4 T_D T}{4 \eta +c^2} (\Gamma -\text{CosIntegral}(2 (\delta/2\pi) T_D )+\log (2 \delta  T_D)), 
\label{infocsbig}
\ee
with $\Gamma$ the Euler constant. For small frequencies, Eq.~\ref{infocsbig} becomes
\be
I_\delta^\mathrm{CS} \approx \frac{2}{\sqrt{\pi}}\frac{c^2\frms^4 T_D^3 T \delta^2}{4 \eta +c^2}.
\ee


In Qdyne, every single phase acquisition period $\tau$ is followed by an interrogation of the NV center state, such that at the end of a measurement the probability to find the NV center on its upper state is $p = \eta + c\sin(\Phi_t)/2$. By keeping $\tau$ constant all through the experiment, and choosing the interrogation time plus NV center reinitialization $\tau_\mathrm{o}$ constant as well, the evolution of $B(t)$ can be probed sequentially every $\tilde\tau = \tau + \tau_\mathrm{o}$. 

If the time $\tilde\tau$ is accurately tracked by, e.g., an external classical clock of sufficient precision (i.e. $\gg T_D$), all measurement outcomes $p$ can be correlated during the postprocessing of the data, such that the autocorrelation now corresponds to the sum of all correlated pairs of measurements, regardless of their temporal spacing, within a single experiment run spanning a total measurement time $T$, composed of $N = T/\tilde\tau$ measurements. The probability to get two correlated measurements separated by an arbitrary time corresponds to the covariance between the measurements, which for a weak signal is $\langle p_0 p_{j\tilde\tau}\rangle \approx \eta^2 + c^2\frms^2 \cos(\delta j\tilde\tau) C(j\tilde\tau / T_D)$. Then, the Fisher information for such a correlated pair is \cite{Oviedo2020}
\be
i_{\delta}^\mathrm{Qd} \approx \frac{c^4}{(4 \eta +c^2)^2}\frms^4 t^2 \sin ^2 (\delta t ) C^2( t/T_\phi ).
\ee
 In this case, the total Fisher information $I_\delta^\mathrm{Qdyne}$ is the sum over all possibly correlated pairs of measurements within an experiment and, contrary to the case of CS, here, for $C(j)\sim j^{-3/2}$, all the terms in $j$ contribute significantly to the total Fisher information. Noting that each measurement has a duration $\tilde\tau$, the total Fisher information reads 
\be
I_\delta^\mathrm{Qdyne} = \frac{c^4 \frms^4}{(4 \eta +c^2)^2}\sum_{j=0}^{T/\tilde\tau}  \left(\frac{T}{\tilde\tau}-j\right)(j\tilde\tau)^2 \sin ^2 (\delta j \tilde\tau ) C^2( j\tilde\tau/T_D),
\ee
with an extra factor $(T/\tilde\tau - j)$ which accounts for all subsequent measurements to which a given measurement can be correlated. As before, we can calculate this sum as an integral, where in addition to using the Riemann rules, we change variables to $z = t/T_D$ for convenience. Then
\be
I_{\delta}^\mathrm{Qd} = \frac{c^4 \frms^4 T_D^4}{(4 \eta +c^2)^2\tilde\tau^2} \int_0^{T/T_D} dz \left(\frac{T}{T_D}-z\right)z^2\sin^2(\delta z T_D) C^2(z).
\ee
Performing the same approximations as with the correlation spectroscopy calculation we get that 
\begin{align}
 I_{\delta}^\mathrm{Qd} &\approx \frac{4}{\sqrt{\pi}}\frac{c^4 \frms^4 T_D^4}{(4 \eta +c^2)^2\tilde\tau^2} \int_0^{T/T_D} dz \left(\frac{T}{T_D}-z\right)z^{-1}\sin(\delta z T_D) \\& \approx \frac{4}{\sqrt{\pi}}\frac{c^4 \frms^4 T_D^4}{(4 \eta +c^2)^2\tilde\tau^2} \frac{-2 \delta  T \text{CosIntegral}(2 T \delta )+\sin (2 \delta  T)+2 \delta  T \left(\log \left(\frac{2 T}{T_D}\right)+\log (\delta  T_D)+\Gamma -1\right)}{4 \delta  T_D}.
 \end{align}
Assuming $(\delta/2\pi) T \gg 1$ and $(\delta/2\pi) T_D < 1$, we get
\be
I_{\delta}^\mathrm{Qd} \approx \frac{2}{\sqrt{\pi}}\frac{c^4 \frms^4 T_D^3 T \log(\delta T)}{(4 \eta +c^2)^2\tilde\tau^2}
\label{eq:FI_Qdyne}
\ee

\subsection{Exponential decay analysis}\label{AppendixF}

Noise originating from the diffusion of molecules in a liquid sample is predominant in statistically polarized nano-NMR, yielding a correlation signal that, at long times, decays as a $3/2$ power law, for which we have demonstrated on the main text that Qdyne is the best protocol to use in most experimental scenarios of relevance. Yet the protocols from quantum sensing have a broad range of applications beyond that of statistically polarized nano-NMR, which means that other noise sources might be dominant, and these are usually modelled as causing an exponential decay of correlations. Thus, in this section, we analyze theoretically the ratio $R_\delta$, between correlation spectroscopy and Qdyne, in the case of noise that causes exponential decay of correlations. Additionally, we analyze the experimental results assuming that the noise model for diffusion is also exponential. While using the incorrect model yields better results than those that would be achieved if the underlying model were truly exponential \cite{Staudenmaier2022}, this permits us showing together theory and experimental results. The analysis procedures are analogous to those presented in the main text. 

In the case of correlation spectroscopy the sum that yields the total Fisher information reads
\be
I_\delta^\mathrm{CS} =  \frac{c^2\frms^4}{4\eta + c^2}\sum_{j=1}^{T/TD} (jT_D)^2 \sin ^2 (\delta j T_D )\exp\left(-2j\right),
\ee
which we calculate by integrating as we did above. In this case, the analytical result can be obtained exactly: 
\begin{equation}
\begin{aligned}
I_\delta^\mathrm{CS} &= \frac{c^2\frms^4T_D T}{4\eta + c^2} \int_0^1 dt\, t^2\sin^2((\delta/2\pi) T_D t)\exp\left(-2t\right) \\&= \frac{c^2\frms^4T_D T}{(4\eta + c^2)8 e^2 \left(\delta ^2 T_D^2+1\right)^3} \big[ \left(e^2-5\right)\left(\delta^6 T_D^6 + 3\delta^4 T_D^4 + 3\delta^2 T_D^2\right) + \left(\delta ^2 T_D^2+5\right) \cos (2 \delta  T_D)  \\&  -\delta  T_D \left(2 \delta ^4 T_D^4+7 \delta ^2 T_D^2+9\right) \sin (2 \delta  T_D)-5 \big] \\&
\approx \frac{3(e^2-7)}{4e^2}\frac{c^2\frms^4\delta^2 T_D^3 T}{(4\eta + c^2)},
\label{csinfoexp}
\end{aligned}
\end{equation}
where the last step involves taking the limit of small frequencies $(\delta/2\pi) T_D < 1$.

For Qdyne, the total Fisher information is, in the case of exponential decay of correlations,
\be
I_\delta^\mathrm{Qdyne} =  \frac{c^4 \frms^4}{(4 \eta +c^2)^2}\sum_{j=0}^{T/\tilde\tau}  \left(\frac{T}{\tilde\tau}-j\right)(j\tilde\tau)^2 \sin ^2 (\delta j \tilde\tau ) \exp( -2j\tilde\tau/T_D),
\ee
which we turn onto an integral that can be calculated exactly:
\begin{equation}
\begin{aligned}
  I_\delta^\mathrm{Qdyne} &= \frac{c^4 \frms^4 T_D^4}{(4 \eta +c^2)^2\tilde\tau^2} \int_0^{T/T_D} dt \left(\frac{T}{T_D}-z\right)z^2\sin(\delta z T_D) \exp(-2z) \\& \approx \frac{c^4 \frms^4 T_D^4}{(4 \eta +c^2)^2\tilde\tau^2} \frac{e^{-\frac{2 T}{T_D}}}{16 T_D^2 \left(\delta ^2 T_D^2+1\right)^4}  \Big[\left(2 T^2+4 T T_D+3 T_D^2\right) \left(\delta ^2 T_D^2+1\right)^4 \\& 
    +4 \delta  T_D \sin (2 \delta  T) \left(\left(\delta ^2 T T_D^2+T\right)^2+T T_D \left(\delta ^4 \left(-T_D^4\right)+2 \delta ^2 T_D^2+3\right)-3 \delta ^2 T_D^4+3 T_D^2\right)\\&
    +\cos (2 \delta  T) \left(2 \left(\delta ^2 T_D^2-1\right) \left(\delta ^2 T T_D^2+T\right)^2+4 T T_D \left(3 \delta ^4 T_D^4+2 \delta ^2 T_D^2-1\right)-3 \left(\delta ^4 T_D^6-6 \delta ^2 T_D^4+T_D^2\right)\right)\\&
    +\delta ^2 T_D^3 e^{\frac{2 T}{T_D}} \left(\delta ^6 T_D^6 (2 T-3 T_D)+4 \delta ^4 T_D^4 (2 T-3 T_D)+3 \delta ^2 T_D^2 (6 T-5 T_D)+12 T-30 T_D\right)\Big],
    \label{qdyneinfoexp}
\end{aligned}
\end{equation}
which, in the limit of low frequencies $(\delta/2\pi) T_D < 1$ reads
\be
I_\delta^\mathrm{Qdyne} \approx \frac{c^4 \frms^4 T_D^2\delta ^2 e^{-\frac{2 T}{T_D}}}{16(4 \eta +c^2)^2\tilde\tau^2} \left(4 T^4+16 T^3 T_D+36 T^2 T_D^2-30 T_D^4 e^{\frac{2 T}{T_D}}+12 T T_D^3 e^{\frac{2 T}{T_D}}+48 T T_D^3+30 T_D^4\right),
\ee
and further applying the limit of large experiment time $(\delta/2\pi) T \gg 1$ and $T\gg T_D$ yields 
\be
I_\delta^\mathrm{Qdyne} \approx \frac{3c^4 \frms^4 T_D^5\delta ^2 T}{4(4 \eta +c^2)^2\tilde\tau^2}
\ee

Then, for exponential correlations, the ratio $R_\delta$ is
\be
R_\delta \approx \frac{e^2c^2T_D^3}{(e^2 - 7)(4\eta + c^2)\tilde\tau^2},
\ee
which, to first order, does not depend on the total experiment time $T$, meaning that Qdyne shows only marginal gains with respect to correlation spectroscopy when the experiment is performed for a long time. The reason being that for exponential correlations, beyond $T_D$, correlations hold no information about the parameters of the signal. This shows that, in this scenario, Qdyne will be superior only for viscous fluids in which many measurements can be performed before the signal decays. In what follows, we explore $R_\delta$ for different parameters and compare with experimental results. 

\begin{figure}
    \centering
   \includegraphics[width=16cm]{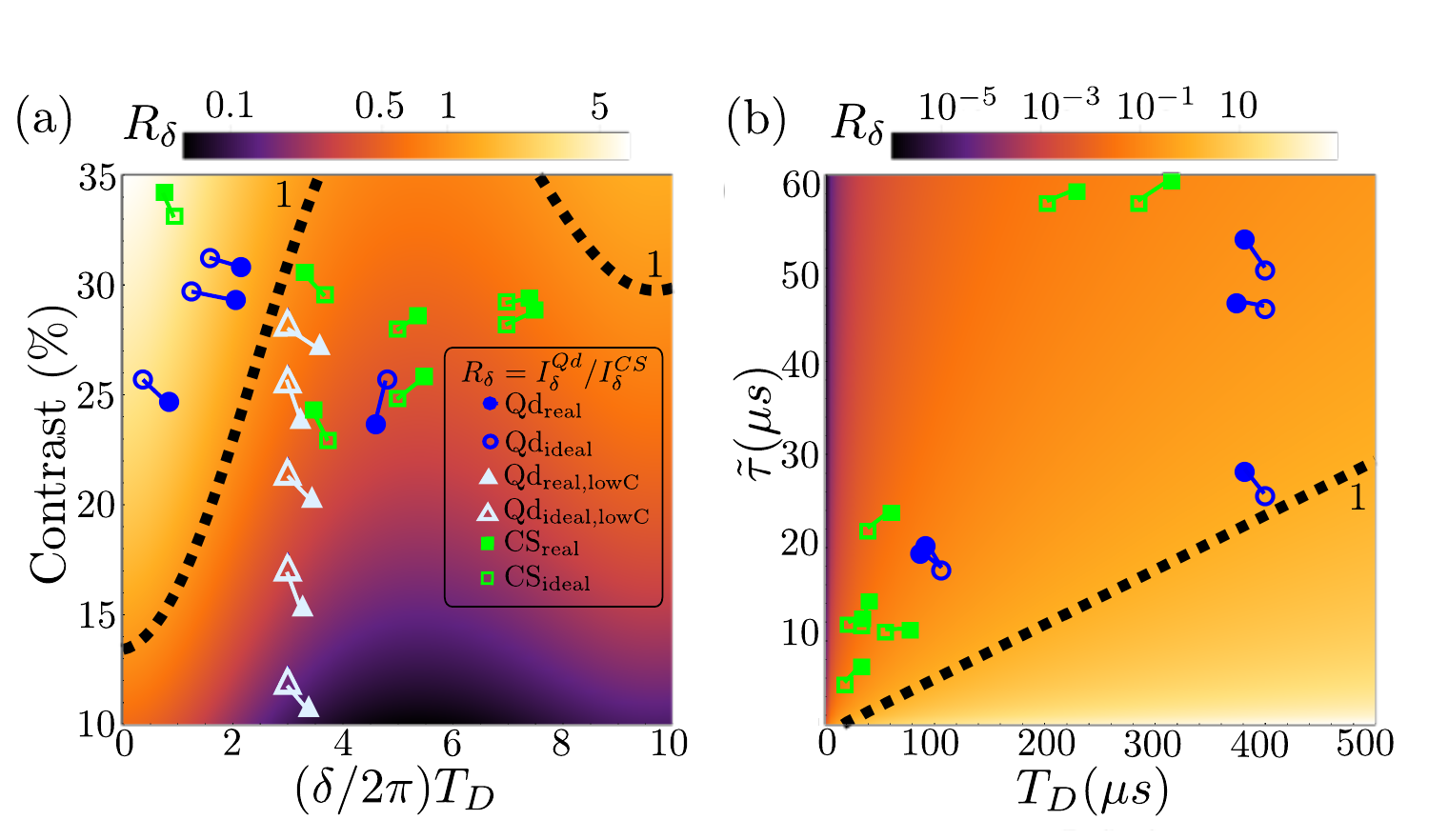}
    \caption{Fisher information ratio $R_\delta$ between experiments and ideal results for exponential correlations. Exact numerical calculation as background with experimental results shown as full shapes for the actual experiments and empty shapes for the position where the experiment would be according to its defining parameters. Thick dashed lines show the reference $R_\delta$ = 1. In (a) we take $T$ = 100 hours, with $T_D$ = 400 $\mu$s and $\tilde\tau$ = 25 $\mu$s, while for (c), we take a contrast $\chi$ = 25 \%, with $(\delta/2\pi) T_D$ = 1 and overhead time $\tau_\mathrm{o}$ = 2.1 $\mu$s.}
    \label{FigExp}
\end{figure}

For each experiment we calculate, as described on the main text, the mean squared error of frequency estimation for a model of autocorrelations with exponential envelope, i.e. $\frms^2\cos(\delta t)\exp{(t/T_D)}$. The inverse of each mean squared error is our measure of the experimental Fisher information. For each experimental parameters we calculate the theoretical Fisher information, from Eqs.~(\ref{csinfoexp}) and (\ref{qdyneinfoexp}), that would correspond to the other protocol, and thus obtain a measure of the ratio $R_\delta$. Fig.~\ref{FigExp} displays these results set in a theoretical background calculated exactly using the Fisher information formulas presented above. To show together theory and experiments, we scale the experimental ratios obtained according to the parameters with which the background theory was calculated. 

In Fig.~\ref{FigExp}(a) we compare the performance of both protocols according to the experimental contrast (defined here as percentage contrast $\chi$), and the relation that the target signal frequency has with the characteristic decay time. We choose a total experiment time of T = 100 hours, and fix $T_D$ to 400 $\mu$s, despite of which we can see that only at very low frequencies does Qdyne offer better resolution capabilities. For shorter decay times, corresponding to fast diffusing fluids, it is not possible to perform enough measurements within a $T_D$ for Qdyne to gather sufficient information and, consequently, correlation spectroscopy is a better strategy. For exponential decay, since correlations are limited to $T_D$, a time-scale beyond which no information about the frequency of the signal can be gathered, the experimental contrast becomes a more important parameter, as evidenced by the huge oscillation that the iso-line marking the boundary $R_\delta$ = 1 shows. 

The sensitivity to the experimental parameters is also shown in Fig.~\ref{FigExp}(b), where for the same experimental time T = 100\,hours, and a fixed contrast of $\chi$ = 25\,\%, we explore the dependence of $R_\delta$ with the correlation time $T_D$ of the sample and the measurement duration $\tilde\tau$. For exponential decay, $T_D$ acts as the limiting time beyond which no information can be acquired from the signal. In this scenario, Qdyne is only advantageous whenever $T_D$ is long, allowing to perform a significant number of measurements before the signal decays, or when $\tilde\tau$ is short, for the very same reasons. For the particular case of diffusion, if we had assumed diffusion to cause exponential decay, $T_D \propto d^{-2}$ and $\Phi_\mathrm{rms} \propto \tau \propto d_\mathrm{nv}^{-3/2}$, meaning that shallow NV centers favor correlation spectroscopy while deeper NVs would benefit from Qdyne. For different noise sources causing exponential decay the particular details might change this conclusion, but not the general one in which only when many measurements can be performed within the correlation time does Qdyne prove superior.

\end{document}